# Extraction of Web Usage Profiles using Simulated Annealing Based Biclustering Approach

R. Rathipriya*, K. Thangavel**

**Abstract**

In this paper, the Simulated Annealing (SA) based biclustering approach is proposed in which SA is used as an optimization tool for biclustering of web usage data to identify the optimal user profile from the given web usage data. Extracted biclusters are consists of correlated users whose usage behaviors are similar across the subset of web pages of a web site where as these users are uncorrelated for remaining pages of a web site. These results are very useful in web personalization so that it communicates better with its users and for making customized prediction. Also useful for providing customized web service too. Experiment was conducted on the real web usage dataset called CTI dataset. Results show that proposed SA based biclustering approach can extract highly correlated user groups from the preprocessed web usage data.

## 1. Introduction

Web personalization is the fundamental task for web recommendation. It is defined as the process of personalizing a web site or web services to the needs of individual web user, by the extracting the knowledge from the analysis of the user's navigational behavior. Due to the rapid growth of the web, the domain of web personalization has gained significant importance in commercial areas. User profiling is the fundamental component of the Web Personalization Systems (WPS). User profiles or Usage models are the generalization of the collected data about the user behavior from web usage data (such as clickstream data). The main goal of user profiling is to increase the efficiency of user activities by delivering more accurate personalized information when users interacting the web site.

Usage-based Personalized Recommendation systems (*Cooley, 1997; Srivatsava, 2000*), analyzed the user's navigation pattern to provide personalized recommendations of web pages according to the current interests of the user. In the literature, various clustering algorithms can be applied to detect the user profiles as well as other web mining techniques such as association rule, have been explored by several researchers. Clustering is the process of grouping the users into clusters such that users within a cluster have high similarity compared to each other but dissimilar to users in other clusters.

Web transactions are grouped into clusters of user sessions or pages by using clustering technique and patterns from the clusters are discovered, by itself is not sufficient for performing the personalization tasks (*Bamshad Mobasher, 2000*).

Clustering is an unsupervised learning technique that attempts to model the trends or patterns within data in order to uncover previously unknown relationships and classifications. Cluster analysis groups objects on the basis of their similarity as determined by a chosen metric which may vary according to context. In the field of the WPS, main limitation of clustering technique is identified such that it partitions users according to their similar browsing interest under all pages of a web site. However, some web users behave similarly only on a subset of pages and their behavior is uncorrelated (dissimilar) over the remaining pages. Therefore, traditional clustering methods will fail to identify such users groups.

* Assistant Professor, Department of Computer Science, Periyar University, Salem,Tamil Nadu, India.
  E-mail: rathi_priyar@yahoo.co.in
** Professor & Head, Department of Computer Science, Periyar University, Salem, Tamil Nadu, India.
  E-mail: drktvelu@yahoo.com

A variation on the clustering approach that supports the discovery of the local browsing patterns within a web usage dataset is that of biclustering. In general, biclustering refers to the 'simultaneous clustering' of both rows and columns of a data matrix [12]. In this paper, Simulated Annealing based Biclustering Approach is used to overcome the above said limitation. In contrast to traditional clustering, a biclustering method produces biclusters, which consists of a subset of users and a subset of pages under which these users behave similarly over the web pages. In literature, Greedy search algorithms are used as the promising approach in the biclustering algorithms. Greedy search algorithms start with an initial solution and find a locally optimal solution by successive transformations that improve some fitness function. Most of the times, it suffers from local optima problem. Meta-heuristics optimization algorithms such as Particle Swarm Optimization (PSO), Genetic Algorithm (GA) and Simulated Annealing (SA) are used along with greedy biclustering to improve the results because it has potential to escape local optima.

The focus of the present study is on the well-known search strategy called Simulated Annealing (SA) optimization technique. SA based biclustering approach is proposed to extract the optimal usage profile/ usage model, which plays a vital role in the recommendation systems. The result of this approach is compared with greedy biclustering approach.

The rest of the paper is organized as follows. Section 2 discussed related work for biclustering methods, and recommendation systems available in the literature. A brief introduction to the preliminaries required for the proposed work is discussed in Section 3. Section 3 describes in detail about the SA based biclustering algorithm for user profiling. The experimental results are discussed in section 5. Section 6 concluded the paper and scope for future enhancement is provided.

## 2. Related Work

Recent research has focused on Web Usage Mining approach for Web Personalization (*Srivatsava, J., 2000*).

The pattern discovery phase, using various data mining techniques, is performed offline to improve the scalability of collaborative filtering. The discovered patterns or aggregate usage profiles can be used to provide dynamic recommendations based on the current user's interest.

In (*AlMurtadha, 2010*), a model has been developed using K-Means clustering approach for deriving usage profiles which followed by recommender systems to predict the next navigation profile. In (*Bamshad Mobasher, 2000*) usage based personalization have been discussed using various data mining techniques. In (Şule Gündüz and M. T. Özsu,2003), model based clustering approach was developed based on the user's interest in a session that are used to recommend pages to the user. WebPersonalizer is the usage based web personalization system to provide dynamic recommendations, was proposed by author of *Bamshad Mobasher, 1999*.

In (*Bamshad Mobasher, 2002*), two different techniques were discussed such as PACT based on the clustering of user transactions and Association Rule Hypergraph Partitioning based on the clustering of pageviews for the extraction of usage profiles. An improved web page prediction accuracy by using a novel approach that involves integrating clustering, association rules and Markov models based on certain constraints has been presented in *Forsat (2009)*. In (*Symeonidis,2006*), use biclustering approach to provide a recommendation to users based on the user and item similarity of neighborhood biclusters.

Sequential Web Access based Recommender System (called SWARS) was proposed in (*Zhou, 2004*) that uses sequential access pattern mining. Liu and Keselj (*Liu, 2007*), proposed an approach that classifies the user navigation patterns and predicting users' future requests based on the combined mining of web server logs and the contents of the retrieved web pages. In (*Castellano et al.,2011*), usage-based web recommendation system is presented that exploits the potential of neuro-fuzzy computational intelligence techniques to suggest interesting pages dynamically to users according to their preferences.

In these works, for obtaining the usage/user profile from web usage data, various forms of clustering techniques are applied very frequently. These clustering algorithms are mainly manipulated on one dimension/attribute of the web usage data only, i.e. user or page solely, rather than taking into account the correlation between web users and pages of a website. However, it failed to extract highly correlated usage groups because highly correlated users group existed only for a subset of pages. In this scenario, web biclustering is probably an effective means to address the mentioned challenge. Therefore, biclustering methods are used in this work to extract hidden users groups based on the subset of pages of a website. It finds a set of significant biclusters in a matrix (i.e.) identify submatrices (subsets of rows and subsets of columns) with interesting properties (*Madeira, 2004*). Perform simultaneous clustering on the row and column dimensions of the matrix instead of clustering the rows and columns separately. Unlike clustering, biclustering identifies groups of users that show similar activity patterns under a specific subset of the pages of a web site. It generates many local patterns. Biclustering is the key technique to use when

- A small set of the users visits the subset set of pages
- An interesting users' browsing pattern is exhibited only in a subset of the pages.

Biclustering methods based optimization algorithms has potential to produce the optimal results than greedy based biclustering method. In our previous *works (Rathipriya et. al, 2011*) Genetic Algorithm and Binary Particle Swarm Optimization are used as an optimization tool to extract the optimal user profile. In this paper, a Simulated Annealing based biclustering approach is proposed to identifying optimal user profiles based on their browsing interest and thereby one can provide a recommendation of web pages effectively. The proposed approach is tested on CTI dataset (Mobasher (2004) and Zhang et al. (2005)). The results indicate that the proposed approach can improve the quality of the aggregate usage profiles.

## 3. Methods and Materials

In this section, preliminaries required for the proposed work are described briefly.

### 3.1 Data Preprocessing

A session file consists of a sequence of user's request for pages P= $\{p_1, p_2, p_3,\ldots,p_n\}$ and a set of m sessions, S = $\{s_1, s_2, s_3,\ldots,s_m\}$ where each $s_i$ belongs to S (*Mobasher, 2004 and Zhang et al., 2005*). A session-pageview matrix A(U, P) of size n x m where n is the number of sessions and m is the number of pageviews. Each row of A represents a user session and each column represents a frequency of occurrence of the pageview in that session.

### 3.2 Evaluation Measure for Bicluster

**A**verage **C**orrelation **V**alue (ACV) was used to evaluate homogeneity of a bicluster in (*Tang C, 2001*). ACV of matrix B($b_{ij}$) is defined by the following function,

$$ACB(B) = \frac{\sum_{i=1}^{n}\sum_{j=1}^{n}|r\_row_{ij}|-n}{n^2-n}, \frac{\sum_{k=1}^{m}\sum_{l=1}^{m}|r\_col_{kl}|-m}{m^2-m} \quad (1)$$

is the correlation between row i and row j and is the correlation between column k and column l. A high ACV suggests high similarities among the rows or columns. ACV can tolerate translation as well as scaling patterns. ACV is robust to noise and it works well for non-perfect bicluster. Other most commonly used bicluster measures are Variance *(Hartigan, 1976)* and Mean Squared Residue (MSR) score *(Cheng et. al, 2000)*, which change dramatically with the magnitude of the bicluster but ACV is more robust for different types of biclusters than the alternatives.

### 3.3 Fitness Function

The fitness function, is normally used to transform the objective function value into a measure of relative fitness.

$$F(x) = g(f(x)) \quad (2)$$

where *f* is the objective function, *g* transforms the value of the objective function to a non-negative number and *F* is the resulting fitness value. The main objective of this work is to find maximal volume biclusters with high ACV. The following fitness function is designed to extract the high volume biclusters from the preprocessed web usage data subjected to ACV threshold δ.
F(B) is defined as

$$F(B) = \begin{cases} |U|*|P|, & if\ ACV(B) \geq \delta \\ 0, & otherwise \end{cases} \quad (3)$$

where |U| and |P| are number of rows and columns of bicluster and δ is defined according to the dataset (i.e taking ACV of entire dataset as a threshold plus some constant between 0 and 1 or set δ greater than 0.9 because ACV of bicluster is 0.9 and above means rows or columns of B are highly correlated).

### 3.4 Representation of Initial Population

SA starts with a group of biclusters known as initial population. The population has N biclusters and is N x L matrix filled with random ones and zeros generated using

population = round(rand( N, L))

where L is the sum of the number of rows and number of columns in the dataset, the function rand(N, L) generates a N x L matrix of uniform random numbers between zero and one. Each row in the matrix represents the single bicluster.

### 3.5 Simulated Annealing (SA): An Overview

Simulated Annealing is a well-established stochastic technique originally developed to model the natural process of crystallization and later adopted to solve optimization problems (*Triki, 2005*). SA is a variant of local neighborhood search. Traditional local search (e.g. steepest descent for minimization) always moves in a direction of improvement whereas SA allows non-improving moves to avoid being stuck at a local optimum.

It has ability to allow the probabilistic acceptance of changes, which lead to worse solutions i.e. reversals in fitness. The probability of accepting a reversal is inversely proportional to the size of the reversal with the acceptance of smaller reversals being more probable. This probability also decreases as the search continues or as the system cools allowing eventual convergence on a solution. It is defined by Boltzman's equation:

$$P(\Delta E) \, \alpha \, e^{\frac{-\Delta E}{T}} \qquad (4)$$

where ∆E is the difference in energy (fitness) between the old and new states and T is the temperature of the system.

In the virtual environment the temperature of the system is lowered after certain predefined number of accepted changes, successes, or total changes, attempts, depending on which is reached first. The rate at which temperature decreases depends on the cooling schedule. In the natural process the system cools logarithmically however this is so time consuming that many simplified cooling schedules have been introduced for practical problem solving; the following simple cooling model is popular:

$$T(k) = \frac{T(k-1)}{(1+\alpha)} \qquad (5)$$

where T(k) is the current temperature, T(k−1) is the previous temperature and α indicates the cooling rate.

Pseudo Code for Simulated Annealing Algorithm

Step 1: Initialize a very high "temperature" and particles.

Step 2: Move the particles to its neighbor.

Step 3: Calculate the fitness of particle.

Step 4: Depending on the change in fitness value, accept or reject the move.

Step 5: Compute probability of acceptance depending on the current "temperature" T using equation (4).

Step 6: Update the temperature value by using equation (5).

Step 6: Go back to Step 2.

Each step of the SA algorithm replaces the current solution by a random nearby solution, chosen with a probability that depends on the difference between the corresponding function values and on a global parameter T called the temperature that is gradually decreased during the process.

### 4. Building the User Profiles from Web Usage Data

The proposed work aimed at to extract highly correlated biclusters from the web usage data. Greedy search algorithm is used to improve the initial solution by finding a locally optimal solution in successive transformations. Stochastic methods such as Simulated Annealing (SA) *(Bryan 2007)* improve on greedy search due to their having the potential to escape local optima. SA based biclustering method that improves on results produced by greedy biclustering algorithm *(Rathipriya et. al. 2011)*. Stochastic techniques that allow acceptance of reversals in fitness have been shown to improve on greedy approaches by performing more in-depth searches of solution space.

**Figure 1:** An Overview of the Proposed Work

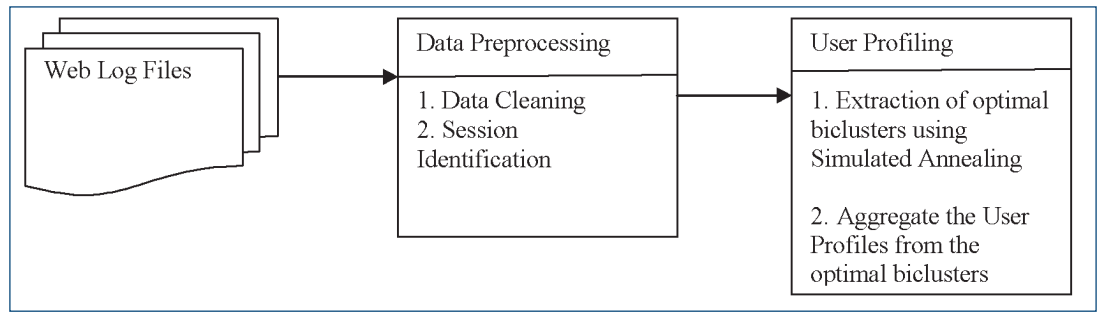

---

Algorithm 1: Simulated Annealing Based Biclustering Algorithm

Input: Session Matrix A, T ,α
Output : Global Optimal Bicluster

Step 1. Initialize k = 0, T, α,Tmin
Step 2. Generate random population.   // Initial Biclusters
         // **extraction of bicluster using SA**
Step 3. initial particles (s) = initial biclusters   // Initialize the initial solution
Step 4. e=fitness(s)
Step 5. ebest=max( e)
Step 6. **while** (T < Tmin)
         snew = generate neighbour of s   // Pick some neighbour.
         2enew = fitness(snew)            // Compute its fitness.
         **if** enew > ebest **then**
                 sbest ← snew
                 ebest ← enew
         **elseif** exp(-(e - enew)/T) >random() then
                 s = snew
                 e= enew
         **else**
         T = T/(1+ α)              // Reduce the temperature
         **End (while)**
Step 7. Return sbest as optimal biclusters

---

Simulated Annealing (SA) based Biclustering algorithm is a top-down biclustering approach. The initial solution i.e. initial population contains all rows (users) and columns (pages) in the web usage dataset. The initial solution is perturbed iteratively by the deletion or addition of web users or web pages whose ACV is calculated for every perturbation. Search algorithms like SA, is used to extract the bicluster from the given dataset it will always produce optimal bicluster without trapping at local optima. This is much less likely to happen with greedy biclustering as it gets trapped in local optima. The critical and significant step in the Web Personalization System (WPS) is the effective derivation of good quality and useful usage or navigation profiles from the web usage dataset. Therefore, SA based biclustering approach is the ideal way to generate user profile from the dataset. Optimal biclusters are used to generate the aggregated user profiles using Algorithm-2.

Algorithm 2: Generation of Aggregate Usage Profile from Bicluster

Input : Set of optimal biclusters, *min_weight*

Output: Set of Aggregate usage profile

　　Step 1.　　For all pages p in global optimal bicluster

　　　　　i. Calculate the weight of the page p using equation (6)

　　　　　ii. If w(p) > *min_weight*

　　　　　　　profile= Union(profile, p)

　　　　　end(if)

　　　end(for)

　　Step 2.　　Return the aggregate user profile

The weight of each page in the bicluster is calculated using following formula as given *(Bamshad Mobasher et al. 2002)*.

$$w(p) = \frac{\sum_{i=1}^{nU} B(i,p)}{nU} \quad (6)$$

where p is the index of the page, 'i' is the user in the bicluster (B) and nU is the number of users in the bicluster. The range of w(p) is 0 to 1.

Aggregated usage profile *(Bamshad Mobasher et al. 2002)* is defined as a set of pairs of pageview and weight.

Aggregated Usage Profile = { p, w(p) | p ∈ P, w(p) ≥ min_weight }

where $P = \{p_1, p_2, \ldots, p_n\}$, a set of *n* pageviews and each pageview uniquely represented by its associated URL and the w(p) is the (mean) value of the attribute's weights in the bicluster.

## 5. Experimental Analysis

### 5.1 CTI Dataset Description

The dataset *CTI* is taken from a university web site log file, which was made available by the authors of Mobasher (2004) and Zhang et al. (2005). The data is based on a random collection of users visiting university site for a 2-week period during the month of April 2002. After data preprocessing, the filtered data consisted of 13745 sessions and 683 pages. CTI dataset is preprocessed again where the root pages were considered in the page view of a session. This preprocessing step resulted in total of 10 categories namely, search, programs, news, admissions, advising, courses, people, research, etc. These page views were given numeric labels as 1 for search, 2 for programs and so on.

Each row of CTI dataset describes the hits of a single user. The session length in the dataset ranges from 2 to 10. Since comparing very long sessions with small sessions would not be meaningful, only sessions of length between 5 and 10 are considered. Finally, 3000 user sessions are taken for the experimentation. Table 1 shows the complete list of numeric coded web pages.

**Table 1:** List of Numeric Coded Web Pages of CTI Dataset

| Search | 1 | Courses | 6 |
|---|---|---|---|
| Programs | 2 | People | 7 |
| News | 3 | Authenticate | 8 |
| Admissions | 4 | CTI | 9 |
| Advising | 5 | Miscellaneous | 10 |

**Table 2:** Parameter Setting for the Proposed Work

| *Parameters* | *Values* |
|---|---|
| N (Number of initial biclusters) | 100 |
| T | 50 |
| A | 0.7 |
| Tmin | 0.01 |
| ACV Threshold δ | 0.93 |

Parameter setting for this study is given in the Table 2. Table 3 tabulates the worst and best solution of both SA and Greedy based biclusters *(Rathipriya et. al, 2011)*. It is evident from the Table 4, that SA based Biclustering has high overlapping degree than greedy biclustering method.

**Table 3: Comparison of SA and Greedy based Biclustering**

|  | ACV of Sbest | | Volume of Sbest | |
|---|---|---|---|---|
|  | Worst Solution | Best Solution | Worst Solution | Best Solution |
| Simulated Annealing Based Biclustering | 0.7019 | 0.9201 | 360 | 4350 |
| Greedy Biclustering | 0.4509 | 0.8903 | 240 | 1699.8 |

**Table 4: Overlapping Degree of the Biclustering Methods**

|  | Overlapping degree |
|---|---|
| Initial Biclusters | 0 |
| Greedy Biclustering | 0.0345 |
| SA based Biclustering | 0.2437 |

The values recorded in the Table 5 show the mean ACV and mean Volume of the biclusters generated by the above two biclustering methods. It is clear that SA based biclustering has high mean ACV than Greedy biclustering method, which is graphically depicted in Figure 2. This improvement is due to the strong local search ability of the Simulated Annealing (SA) techniques.

**Table 5: Performance of the Biclustering Methods**

|  | Greedy Biclustering | SA based Biclustering |
|---|---|---|
| Mean Volume | 2345 | 7365 |
| Mean ACV | 0.8976 | 0.9289 |

From the extracted biclusters, aggregate usage profile is generated using Algorithm- 2 and it is shown in the Table-6 tabulates the list pages and weights of each page in the aggregate usage profile. The percentage of the users in each optimal bicluster is given in the fifth column. The first user profile has highest percentage of users and high ACV. These usage profiles served as model to predict the set of web pages for next navigation based on their current navigation profiles.

**Figure 2: Comparison of Biclustering Methods**

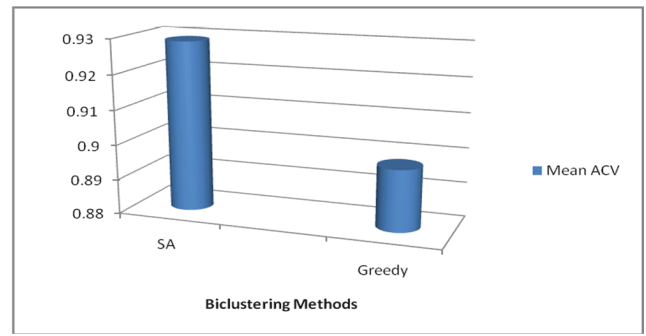

**Table 6: Aggregate Usage Profiles Using SA Based Biclustering**

| Aggregate Usage Profile | List of Pages | Weights of each page in the Profile | ACV | Percentage of Users in Bicluster |
|---|---|---|---|---|
| 1 | 1,3,5,6,7,10 | 0.6790 , 0.7819, 0.6533, 0.7295, 0.7048, 0.6190 | 0.9452 | 76.56% |
| 2 | 1,4,7 | 0.7421, 1.0000, 0.8507 | 0.9348 | 71.03% |
| 3 | 2,6,8,10 | 1.0000, 0.6804, 0.6041, 0.6701 | 0.9856 | 65.62% |

**Table 7: Pageviews in the Optimal User Profile**

| Optimal User Profile | Weights of Pageview |
|---|---|
| Search | 0.679 |
| News | 0.7819 |
| Advising | 0.6533 |
| Courses | 0.7295 |
| People | 0.7048 |
| Miscellaneous | 0.619 |

Optimal biclusters are post processed to create the user profile as mentioned in the Algorithm-2, which contains a set of pageviews with weights. These weights signify the importance of the pageview in that user profile. Based on these user profiles, the web page is personalized. List of the Pageviews in the highly correlated user profile is tabulated in the Table 7. Almost all pageviews in it have equal weight, this is due their correlated browsing behavior.

## 6. Conclusion

This paper proposed SA based Biclustering approach for web user profiling. SA has the potential to give improved results for the biclustering problem. From this study, it is concluded that proposed biclustering approach performs better than greedy biclustering approach. This improvement is due to strong search ability of SA. It discovers highly significant user profiles, which play a vital role in the web personalization systems. Biclustering results are analyzed and could be useful in a great number of applications such as e-commerce applications, recommendation engines, system performance, website customization, etc,. A future work is to improve the efficiency of the proposed biclustering approach by adopting different methods to generate the neighbor solution in SA for fast convergence. Greedy search algorithms are used as the promising approach in the biclustering algorithms. Greedy search algorithms start with an initial solution and find a locally optimal solution by successive transformations that improve some fitness function. Most of the times, its biclustering results suffers from local optima problem. Metaheuristics optimization algorithms such as Particle Swarm Optimization (PSO), Genetic Algorithm (GA) and Simulated Annealing (SA) are used along with greedy biclustering to improve the results because it has potential to escape local optima.

## 7. Acknowledgement

The second author acknowledges the UGC, New Delhi for financial assistance under minor research project under grant no. 41-1354/2012(SR)